\pdfoutput=1
\documentclass{agujournal2019}
\usepackage[intlimits]{amsmath}
\usepackage{amssymb}
\usepackage[
  active=false,
  copydocumentclass=false,
  extract-env=figure,
  generate=figs_sep,
  handles=false,
]{extract}
\journalname{Earth and Space Science}

\begin{document}

%%%%%%%%%%%%%%%%%%%%%%%%%%%%%%%%%%%%%%%%%%%%%%%
%  TITLE
%
% (A title should be specific, informative, and brief. Use
% abbreviations only if they are defined in the abstract. Titles that
% start with general keywords then specific terms are optimized in
% searches)
%
%%%%%%%%%%%%%%%%%%%%%%%%%%%%%%%%%%%%%%%%%%%%%%%

% Example: \title{This is a test title}

\title{Empirical model of SSUSI-derived auroral ionization rates}

%%%%%%%%%%%%%%%%%%%%%%%%%%%%%%%%%%%%%%%%%%%%%%%
%
%  AUTHORS AND AFFILIATIONS
%
%%%%%%%%%%%%%%%%%%%%%%%%%%%%%%%%%%%%%%%%%%%%%%%

% Authors are individuals who have significantly contributed to the
% research and preparation of the article. Group authors are allowed, if
% each author in the group is separately identified in an appendix.)

% List authors by first name or initial followed by last name and
% separated by commas. Use \affil{} to number affiliations, and
% \thanks{} for author notes.
% Additional author notes should be indicated with \thanks{} (for
% example, for current addresses).

% Example: \authors{A. B. Author\affil{1}\thanks{Current address, Antartica}, B. C. Author\affil{2,3}, and D. E.
% Author\affil{3,4}\thanks{Also funded by Monsanto.}}

\authors{Stefan Bender\affil{1,2,3}, Patrick J. Espy\affil{2,3}, Larry J. Paxton\affil{4}}

\affiliation{1}{Instituto de Astrof\'isica de Andaluc\'ia (CSIC), Granada, Spain}
\affiliation{2}{Department of Physics, Norwegian University of Science and Technology, Trondheim, Norway}
\affiliation{3}{Birkeland Centre for Space Science, Bergen, Norway}
\affiliation{4}{Applied Physics Laboratory, Johns Hopkins University, Laurel, Maryland, USA}
%(repeat as many times as is necessary)

% Corresponding author mailing address and e-mail address:

% (include name and email addresses of the corresponding author.  More
% than one corresponding author is allowed in this LaTeX file and for
% publication; but only one corresponding author is allowed in our
% editorial system.)

% Example: \correspondingauthor{First and Last Name}{email@address.edu}

\correspondingauthor{Stefan Bender}{sbender@iaa.es}

%%%%%%%%%%%%%%%%%%%%%%%%%%%%%%%%%%%%%%%%%%%%%%%
% KEY POINTS
%%%%%%%%%%%%%%%%%%%%%%%%%%%%%%%%%%%%%%%%%%%%%%%
%  List up to three key points (at least one is required)
%  Key Points summarize the main points and conclusions of the article
%  Each must be 140 characters or fewer with no special characters or punctuation and must be complete sentences

% Example:
% \begin{keypoints}
% \item	List up to three key points (at least one is required)
% \item	Key Points summarize the main points and conclusions of the article
% \item	Each must be 140 characters or fewer with no special characters or punctuation and must be complete sentences
% \end{keypoints}

\begin{keypoints}
\item We present an empirical model for auroral ionization rates between 90 and 150~km derived from SSUSI FUV observations.
\item The model is based on linear regression with respect to the geomagnetic Kp, PC, and Ap indices,
  and the solar F10.7 radio flux.
\item We find that Kp captures the largest fraction of the ionization rate variability.
\end{keypoints}

%%%%%%%%%%%%%%%%%%%%%%%%%%%%%%%%%%%%%%%%%%%%%%%
%
%  ABSTRACT and PLAIN LANGUAGE SUMMARY
%
% A good Abstract will begin with a short description of the problem
% being addressed, briefly describe the new data or analyses, then
% briefly states the main conclusion(s) and how they are supported and
% uncertainties.

% The Plain Language Summary should be written for a broad audience,
% including journalists and the science-interested public, that will not have 
% a background in your field.
%
% A Plain Language Summary is required in GRL, JGR: Planets, JGR: Biogeosciences,
% JGR: Oceans, G-Cubed, Reviews of Geophysics, and JAMES.
% see http://sharingscience.agu.org/creating-plain-language-summary/)
%
%%%%%%%%%%%%%%%%%%%%%%%%%%%%%%%%%%%%%%%%%%%%%%%

%% \begin{abstract} starts the second page

\begin{abstract}
  We present an empirical model for auroral (90--150\,km)
  electron--ion pair production rates,
  ionization rates for short,
  derived from SSUSI
  (Special Sensor Ultraviolet Spectrographic Imager)
  electron energy and flux data.
  Using the~\citeA{Fang2010} parametrization for mono-energetic electrons,
  and the NRLMSISE-00 neutral atmosphere model~\cite{Picone2002/12/24},
  the calculated ionization rate profiles are binned in
  2-h magnetic local time (MLT) and 3.6$^{\circ}$ geomagnetic latitude
  to yield time series of ionization rates at 5-km altitude steps.
  We fit each of these time series to the geomagnetic indices Kp, PC, and Ap,
  the 81-day averaged solar F$_{\text{10.7}}$ radio flux index, and a constant term.
  % some results?
  The resulting empirical model can easily be incorporated into coupled
  chemistry--climate models to include particle precipitation effects.
\end{abstract}

\section*{Plain Language Summary}
% \note[Wiley]{%
% Enter your Plain Language Summary here or delete this section.
% Here are instructions on writing a Plain Language Summary: 
% \url{https://www.agu.org/Share-and-Advocate/Share/Community/Plain-language-summary}
% }
Aurorae or polar lights are produced at around 100\,km at high latitudes
by electrons and protons from space that enter the upper atmosphere,
approximately around 65$^{\circ}$ North and South.
Besides creating beautiful auroral displays, these particles
also change the atmospheric composition by ionizing the air and thus
initiating chemical reactions.
Chemistry--climate models have recently started implementing these changes
in their simulations.
However, so far these simulations do not match observations.
One possible reason is that the input for the models is based on satellite observations
of particles far above 100\,km, without knowing exactly how many
of them enter the atmosphere.
Here we present a way to use space observations of the aurora to
calculate the number of particles actually entering the atmosphere.
We use these aurora observations to construct a mathematical formula to
calculate how much of the atmosphere is ionized by the electrons and protons.
Our formula is based on indices of geomagnetic activity, and it
can be used in more complicated chemistry--climate models
to better incorporate these effects in their simulations.

%%%%%%%%%%%%%%%%%%%%%%%%%%%%%%%%%%%%%%%%%%%%%%%
%
%  BODY TEXT
%
%%%%%%%%%%%%%%%%%%%%%%%%%%%%%%%%%%%%%%%%%%%%%%%

%%% Suggested section heads:
\section{Introduction}\label{sec:intro}

Particle precipitation and the associated processes initiated in the
middle and upper atmosphere have been
recognized as one component of natural climate
variability and are included in the most recent climate prediction
simulations initiated by the
Intergovernmental Panel on Climate Change (IPCC)~\cite{Matthes2017}.
So far, most of the model inputs are based on in-situ
particle observations at satellite orbital altitudes
($\approx$800\,km)~\cite<e.g.>{Wissing2009, Kamp2016, Smith-Johnsen2018a},
or on trace-gas observations~\cite{Randall2009, Funke2017}.
In addition, most recent studies focus on the influence
of ``medium-energy'' electrons (MEE)
(30--1000\,keV)~\cite{Smith-Johnsen2018a, Tyssoey2021, Sinnhuber2021}
that have their largest impact in the mesosphere ($\lessapprox$90\,km).
But all three of these studies have \emph{also} found a considerable
discrepancy between observations and models regarding
the nitric oxide (NO) content in the \emph{lower thermosphere} (100--120\,km).
In the latest studies by~\citeA{Tyssoey2021} and~\citeA{Sinnhuber2021},
this part of the atmosphere is modelled using
total energy flux and an auroral-oval parametrization based on Kp,
or using
ionization rates provided by the Atmosphere Ionization Module Osnabrück (AIMOS;~\citeA{Wissing2009}),
which is derived from aforementioned in-situ particle observations.
However, inferring the flux precipitating into the atmosphere from the in-situ
particle measurements has its own difficulties and uncertainties,
for example determining the loss cone~\cite{Tyssoey2016}.

In addition to the direct ionization and chemical impact of MEE
on the mesosphere, the descent of NOx from the lower thermosphere
to the stratosphere has been observed~\cite{Randall2009, Funke2005b, Funke2014}.
This transport happens regularly during the final warming and the
breakup of the polar vortex in late winter and spring,
and during Sudden Stratospheric Warmings with Elevated-Stratopause
events~\cite{Perot2014, Perot2021}.
This indirect source of NOx in the middle atmosphere has its origin in the
auroral production of NO in the lower thermosphere.
So far this auroral NO source is mostly described by simple parametrizations
in whole-atmosphere chemistry-climate models~\cite{Sinnhuber2021},
with the exception of AIMOS which is derived from POES particle flux measurements~\cite{Wissing2009}.
Although the model comparison study by~\citeA{Sinnhuber2021}
found good agreement of auroral NO with data \emph{on average},
the amounts differ by an order of magnitude between the individual models.
Our model adds another approach to modelling this auroral NO source in
global climate models, derived from direct auroral UV emission observations.

The Special Sensor Ultraviolet Spectrographic Imagers (SSUSI)
are UV imagers (115--180\,nm) on board the
Defense Meteorological Satellite Program (DMSP) Block-5D3
satellites F17 and F18~\cite{Paxton1992, Paxton2017, Paxton2018}.
These instruments picture
approximately 3000\,km wide swaths of both auroral zones
with a pixel resolution of 10$\times$10\,km at the nadir point.
By observing the atmospheric emissions directly,
this avoids the problem of having to model
the input into the atmosphere as is the case with
the aforementioned in-situ particle measurements.

From these data we
derived ionization rate profiles~\cite{Bender2021},
and here we develop an empirical regression model of these rates
binned in magnetic local time (MLT) and geomagnetic latitude.
The model uses 4 geomagnetic and solar activity indicators:
the geomagnetic Kp and Ap indices,
the auroral Polar Cap (PC) index,
and the 81-day centred moving average of the solar 10.7-cm radio flux.
Our model complements the existing ionization rate models for higher-energetic electrons
such as ApEEP~\cite{Kamp2016, Kamp2018};
it is also based on geomagnetic indices and ready to use in
chemistry--climate models that include the upper atmosphere.
The intended main purpose of our model is to drive the auroral ionization input
in the lower thermosphere in whole-atmosphere chemistry-climate models.
Furthermore, the presented model provides the uncertainties in the proxy coefficients
which can be used to estimate the uncertainty in the modelled
ionization rates.
Those can be further used to carry out statistical ensemble runs of
chemistry--climate models, randomly driven by the ionization rates within their
uncertainties.

The paper is organized as follows:
the data and processing steps are described in Sect.~\ref{sec:data},
the model setup is laid out in Sect.~\ref{sec:model},
and the results are presented in Sect.~\ref{sec:results}.

\section{Data}\label{sec:data}

We use the electron energy and energy flux data from the
Special Sensor Ultraviolet Spectrographic Imager (SSUSI)
instruments on board the Defense Meteorological Satellite Program (DMSP)
Block-5D3 satellites F17 and F18~\cite{Paxton1992, Paxton2017, Paxton2018}.
These satellites orbit at 850\,km altitude in polar,
sun-synchronous orbits,
the equator crossing times of their ascending nodes are 17:34 LT (F17) and 20:00 LT (F18).
The orbital period is of the order of 100\,min,
resulting in about 15 orbits per day.
These instruments provide about 3000\,km-wide
spectrographic images of the auroral zones
with a 10\,km$\times$10\,km pixel size at the nadir point.
For bandwidth reasons, they downlink 5 UV channels including
two colours for the LBH (Lyman--Birge--Hopfield) band emissions of N$_2$.
From these two LBH colours the electron energies and energy fluxes
are derived based on~\citeA{Strickland1983},
and an extensive discussion can be found in~\citeA{Knight2018}.
The electron energy and flux data we use here are provided within the
Auroral-EDR (Environmental Data Record) data set at
\url{https://ssusi.jhuapl.edu/data_products}~\cite{SSUSI2020}.

For each pixel we calculate the ionization rate (IR) profile as described
in~\cite{Bender2021} for altitudes 90--150\,km.
Briefly, we use the parametrization by~\cite{Fang2010} for mono-energetic electrons
and integrate over a Maxwellian spectrum for magnetic local times (MLT) $\leq$12:00
and over a Gaussian spectrum for MLT$>$12:00 as determined by our comparison
study~\cite{Bender2021}.
We use NRLMSISE-00~\cite{Picone2002/12/24} for the background neutral atmosphere,
calculated at each pixel location.
By converting the ionization rates to electron densities,
these data have previously been validated against EISCAT ground-based
measurements~\cite{Bender2021}.

\section{Model description}\label{sec:model}

We bin the ionization rate profiles from the SSUSI data
in 2-h magnetic local time (MLT) bins
from 00:00--02:00 MLT to 22:00--00:00(+1\,d) MLT and in
3.6$^{\circ}$ geomagnetic latitude bins from 52.2$^{\circ}$ (50.4--54.0$^{\circ}$)
to 88.2$^{\circ}$ (86.4--90.0$^{\circ}$) in both hemispheres.
In each bin the ionization rate profiles are calculated using
the parametrization by~\citeA{Fang2010} on a 5-km altitude grid
from 90 to 150\,km, evaluated directly for the given altitude $z$.
The spectra are used according to the validation study by~\citeA{Bender2021}.
We chose to average the profiles instead of the energy and fluxes
because converting them to ionization rates is highly non-linear
in electron energy~\cite{Fang2010}.
This approach has the advantage that mixed spectra are better represented
in the final ionization input which are hard to capture with average parameters.
For example the case of a mixture of two Maxwellian-type spectra as
presented in~\citeA[Fig.\ 4]{Fang2010};
the average has no analytical representation and even using a
different kind of spectra is unable to reproduce the mixed spectrum fully.
As a result, the alternative approaches fail to reproduce the
``double-peak'' structure of the ionization rate profile,
which can be retained to some extend by using the average profile
instead of the average parameters.

In each bin the vertical profiles are averaged per orbit,
yielding a time series of ionization rates $q$ with a standard deviation
for every 5-km altitude step.
To reduce the low bias introduced by orbits with very
few non-zero ionization rate data points in the respective grid cell,
we omit the data from those orbits where the average $q$ is smaller
than 0.001 times the median of the time series.
We use these time series in each bin to fit a multi-linear
regression model to $\log(q)$\footnote{%
  Unless stated otherwise, $\log()$ denotes the natural logarithm
  as the inverse of the exponential $\exp()$
  which is defined by the solution of
  $\frac{\text{d}}{\text{d}x} \exp(x) = \exp(x)$ with $\exp(0) = 1$.
} as follows:
\begin{linenomath*}
\begin{equation}\label{eq:regressmodel}
  \log(q) =
    a \cdot \text{Kp}
    + b \cdot \text{PC}
    + c \cdot \text{Ap}
    + d \cdot \log(\overline{\text{F}_\text{10.7}})
    + \text{const.}
    + \varepsilon
    \;.
\end{equation}
\end{linenomath*}
In Eq.~\eqref{eq:regressmodel},
Kp and Ap are the 3-h geomagnetic indices~\cite<e.g.>{Mandea2011},
PC is the 1-h polar cap index~\cite{Troshichev1979, Troshichev1988},
$\overline{\text{F}_\text{10.7}}$ is the 81-day moving average of
the solar 10.7\,cm radio flux, centred on the day to be fitted.
This list of proxies has been empirically determined by iterating the
available options and where feasible their natural logarithm,
adding one by one and keeping those with the best
fit in terms of maximum likelihood.
The iteration was stopped when further improvement
in terms of the ``Bayesian Information Criterion''~\cite<see, e.g.>{Wit2012}
did not justify adding another regressor to the list.
That is, the proxies in Eq.~\eqref{eq:regressmodel}
are ordered by (empirical) relevance.
All the indices are taken from the OMNI space weather index
database~\cite{King2005} and hourly-sampled values are used~\cite{Papitashvili2020}.
We use Bayesian linear regression with conjugate priors to fit the model coefficients.
We use a wide normal distribution around zero ($\sigma = 10$)
as the prior distribution for the coefficients,
and a wide inverse Gamma distribution ($\alpha = 1, \beta = 1$)
as the prior distribution for the variance.

Note that because of the F17 and F18 orbits,
not all latitude/MLT bins are sampled equally.
Therefore, in bins with low number of points,
that is with less than 240 data points,
we use bilinear interpolation in MLT and geomagnetic latitude
to calculate the coefficient.

\section{Results}\label{sec:results}

\subsection{Time series fit}\label{sec:fit.NH}

We demonstrate the fit quality on an example time series of
the model bin centred at 19:00 MLT (18:00--20:00 MLT),
70.2$^{\circ}$N geomagnetic latitude (68.4--72.0$^{\circ}$N),
and at 100\,km altitude.
This constitutes an example where we have almost
60000 data points available,
and it served as a testbed for determining the proxies
used as regressors in the model.
Note that because of the large range of ionization rates
from $<1$ to $10^6$\,cm$^{-3}$\,s$^{-1}$,
the better choice for fitting is $\log(q)$.

\begin{figure}
\noindent\includegraphics[width=0.98\textwidth]{%
  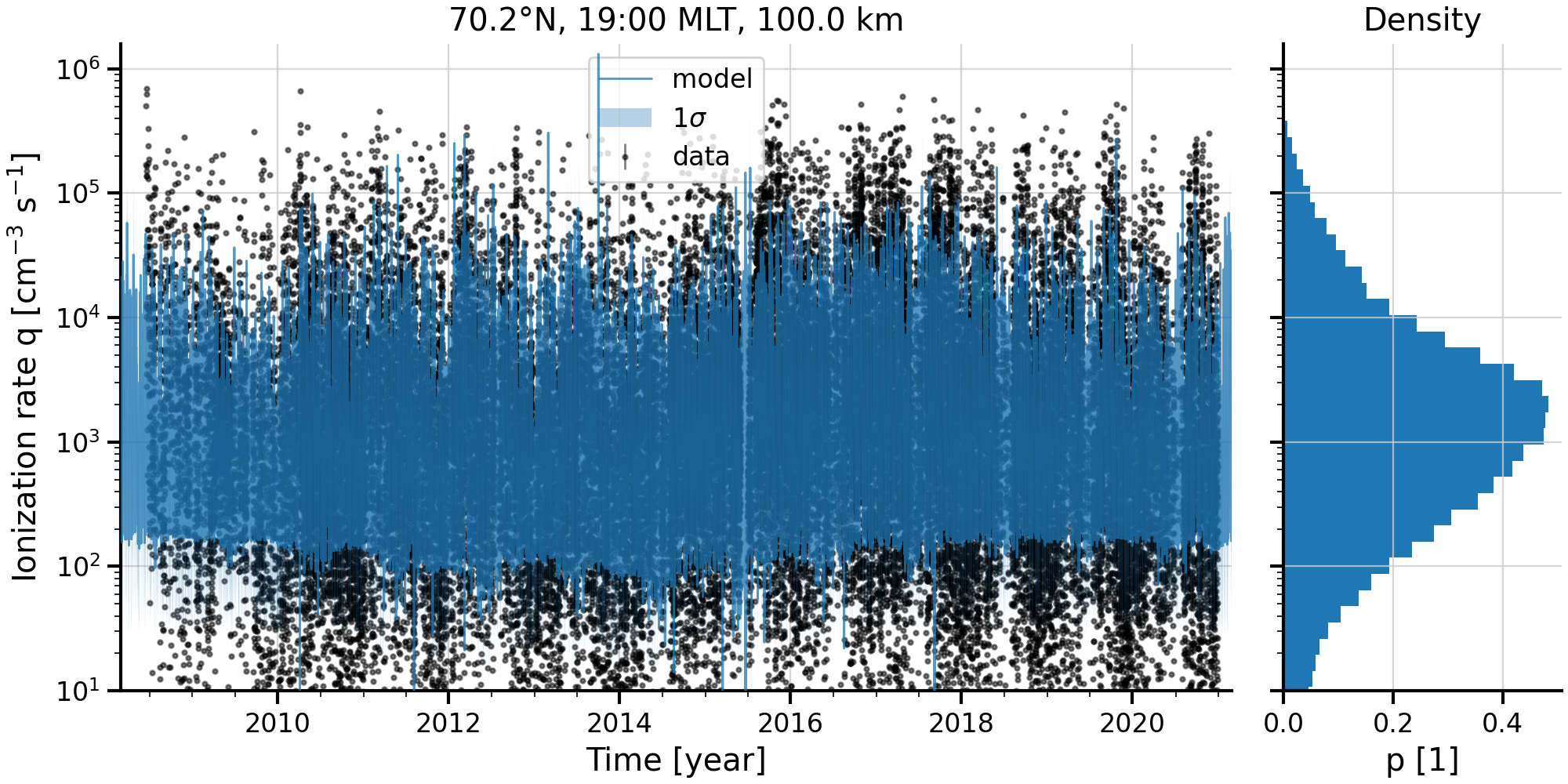%
}
\caption{Data and model fit for example bin:
  19:00 MLT,
  70.2$^{\circ}$N geomagnetic latitude,
  100\,km altitude.
  The histogram shows the distribution of the data.
}
\label{fig:fit.ts.NH}
\end{figure}

\begin{figure}
\noindent\includegraphics[width=0.98\textwidth]{%
  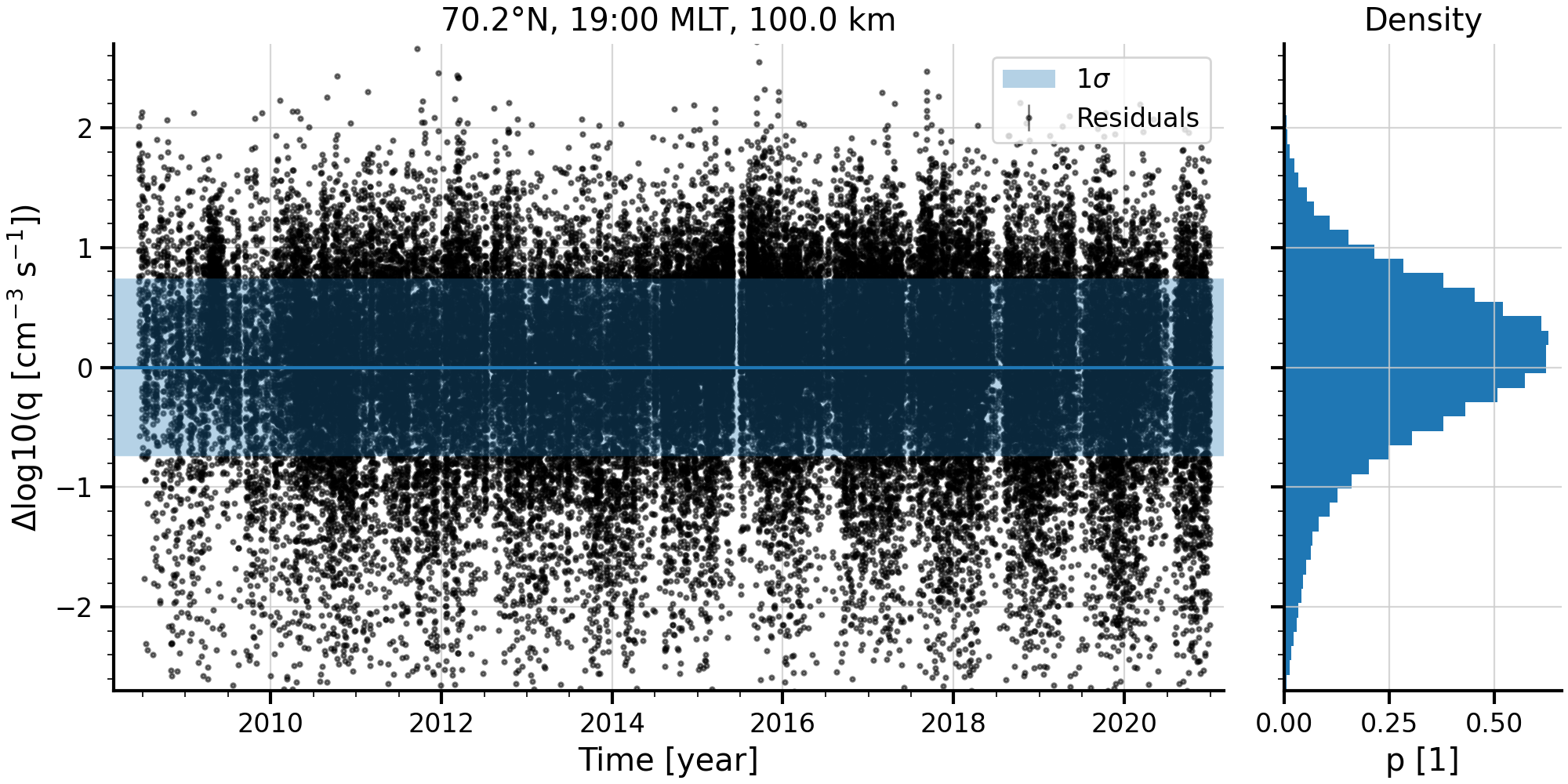%
}
\caption{Residuals for example bin:
  19:00 MLT,
  70.2$^{\circ}$N geomagnetic latitude,
  100\,km altitude.
  The histogram shows the distribution of the residuals.
}
\label{fig:res.ts.NH}
\end{figure}

The data in that bin, together with the model fit, are shown
in Fig.~\ref{fig:fit.ts.NH}.
The residuals and the 1$\sigma$ prediction uncertainty are shown
in Fig.~\ref{fig:res.ts.NH}.
The overall average ionization rate is around $10^{3}$\,cm$^{-3}$\,s$^{-1}$.
The fit line indicates a slight variation following approximately an 11-year solar cycle.
On top of that, annual and semi-annual variations are visible in the data.
Although we do not explicitly include (semi)annual harmonics
in the model, they seem to be captured by
the combination of the proxies used.
Both histograms in Figs.~\ref{fig:fit.ts.NH} and~\ref{fig:res.ts.NH},
showing $\log(q)$ and the residuals, show smooth, almost symmetric distributions.
Thus, the assumption of normal-distributed errors for multi-linear regression is satisfied.

\subsection{Coefficient distributions in the Northern Hemisphere}\label{sec:coeff.NH}

\begin{figure}
\centering
\noindent\includegraphics[width=0.96\textwidth]{%
  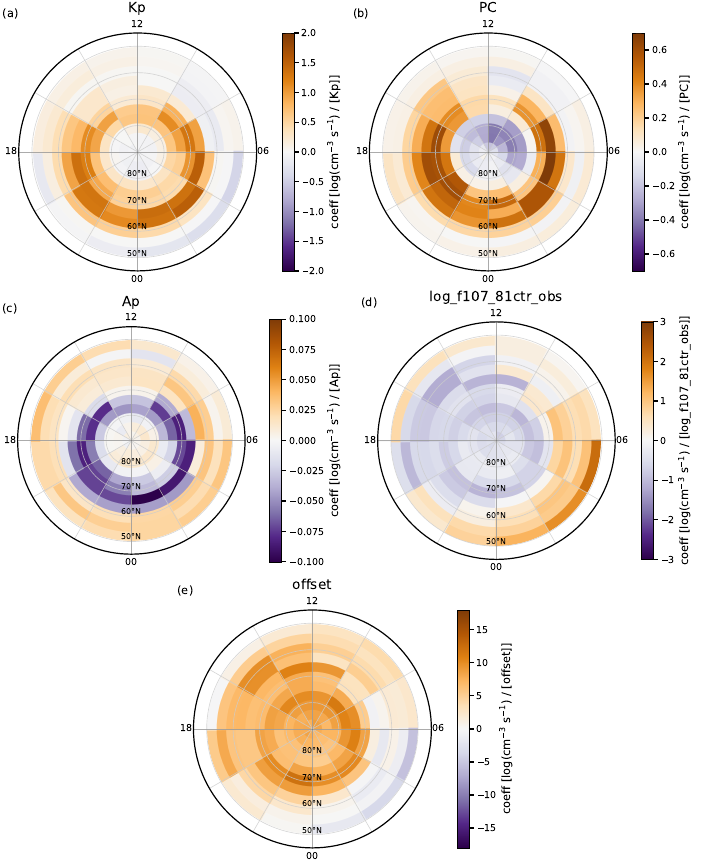%
}
\caption{%
  Fitted regression coefficients in the NH per MLT and geomagnetic latitude,
  (a) Kp, (b) PC, (c) Ap, (d) $\log(\overline{\text{F}_\text{10.7}})$,
  and (e) offset.
}
\label{fig:coeff.NH.KpPC}
\end{figure}

The fit coefficients in the Northern Hemisphere
for 100\,km altitude are shown in Fig.~\ref{fig:coeff.NH.KpPC}.
The coefficients for Kp and PC show similar patterns,
both have their largest effect in a wide band around
auroral geomagnetic latitudes,
between about 55$^{\circ}$ and 75$^{\circ}$ with a maximum around 65$^{\circ}$.
This similarity suggests that PC accounts for shorter variations
(1-h) on top of the 3-h variations captured by Kp,
and essentially both describe the same processes.
We also observe that the Ap coefficients exhibit a similar pattern,
albeit with a negative sign compared to Kp and PC.
This may be a sign that the quasi-logarithmic scale of Kp
introduces larger changes that are compensated by the linear scale of Ap.
The circular patterns in Kp and Ap closer to the pole
for MLT 06:00--18:00
are most likely related to cusp precipitation~\cite{Newell2005, Newell2009}.
Over almost the entire polar cap, the F10.7 coefficients are distributed
opposite to the constant offset coefficients.
This indicates that the influence of the solar cycle is marginal,
but it reduced the fitting metric just enough to warrant its inclusion.

\subsection{Coefficient distributions in the Southern Hemisphere}\label{sec:coeff.SH}

\begin{figure}
\centering
\noindent\includegraphics[width=0.96\textwidth]{%
  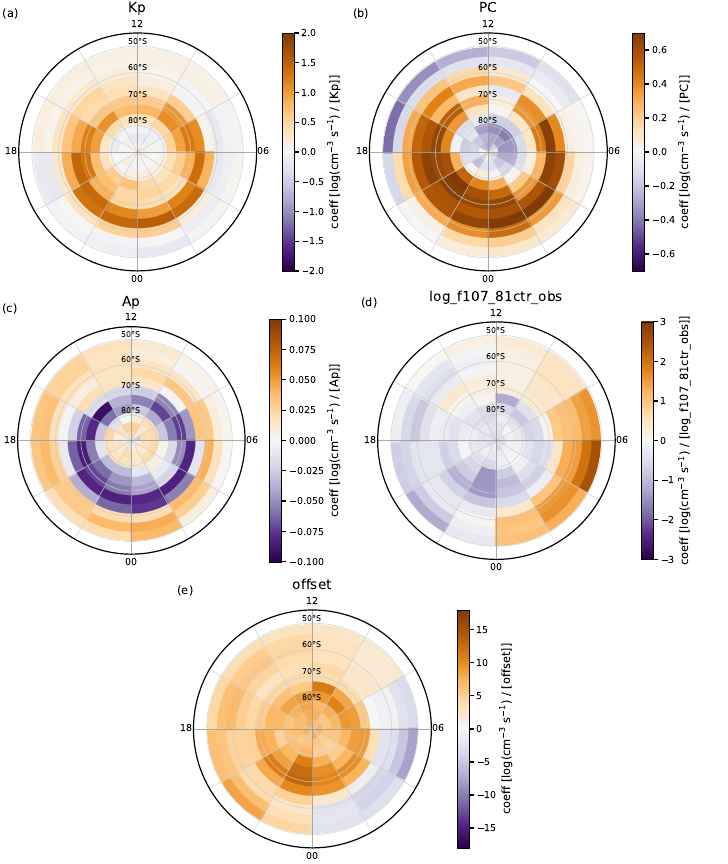%
}
\caption{%
  Fitted regression coefficients in the SH per MLT and geomagnetic latitude,
  (a) Kp, (b) PC, (c) Ap, (d) $\log(\overline{\text{F}_\text{10.7}})$,
  and (e) offset.
}
\label{fig:coeff.SH.KpPC}
\end{figure}

The fit coefficients in the Southern Hemisphere
for 100\,km altitude are shown in Fig.~\ref{fig:coeff.SH.KpPC}.
The coefficients in the Southern Hemisphere exhibit essentially the same patterns
as in the Northern Hemisphere.
The coefficients of the geomagnetic indices are of the same
magnitude as in the Northern Hemisphere,
with slight differences in the MLT/latitude distributions.
For PC the extent of positive coefficients is larger in latitude
than in the Northern Hemisphere.
These small differences could be a sign that the precipitation is not
fully symmetric in geomagnetic latitude between North and South,
and may also be the result of the geomagnetic indices being almost exclusively derived
from observations in the Northern Hemisphere.

\begin{figure}
\centering
\noindent\includegraphics[width=0.96\textwidth]{%
  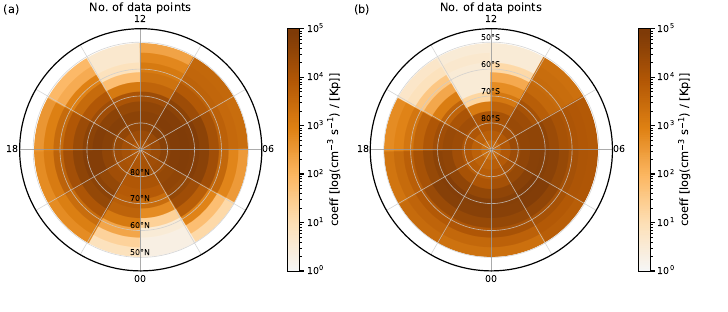%
}
\caption{%
  Distribution of available data points used for fitting, (a) Northern Hemisphere,
  (b) Southern Hemisphere.
}
\label{fig:coeff.npts}
\end{figure}

\subsection{Profile comparison to data and AISstorm}\label{sec:prof.comp}

\begin{figure}
\includegraphics[width=0.98\textwidth]{%
  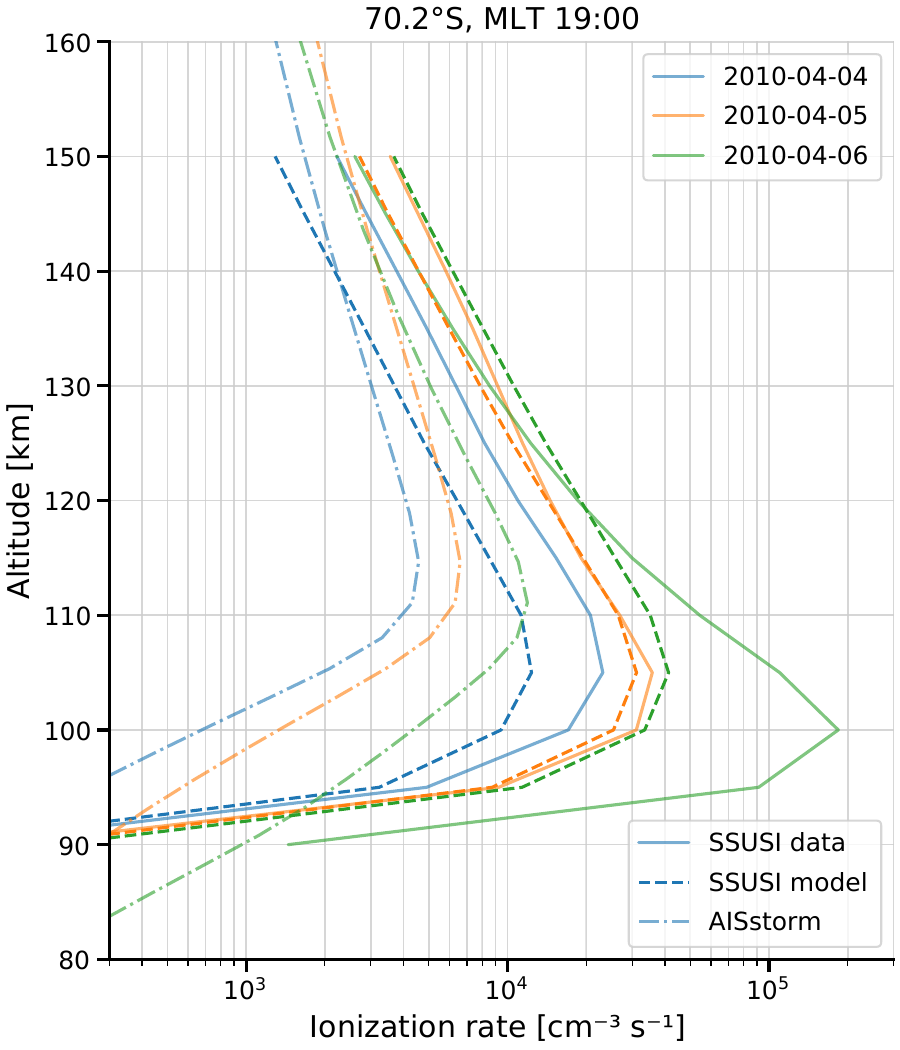%
}
\caption{%
  Daily-averaged ionization rate profiles during a substorm event
  in April 2010~\cite{Tyssoey2021, Sinnhuber2021} for a selected
  geomagentic latitude/MLT bin.
  Shown are the SSUSI-derived profiles (solid),
  the profiles from the model (current work; dashed),
  and the preliminary AISstorm profiles (dash-dotted).
}
\label{fig:profs}
\end{figure}

The overall performance of the model is difficult to assess
based on individual profiles because of its average nature.
As shown for example in Fig.~\ref{fig:res.ts.NH},
the residual variation is still large and for single days,
the modelled ionization rates can differ considerably from the data.
In Fig.~\ref{fig:profs}, we compare the modelled ionization rate profiles to
the original data during the event in April 2010 investigated
in previous studies~\cite{Tyssoey2021, Sinnhuber2021}.
For this comparison we picked a single geomagnetic latitude/MLT bin,
corresponding to 70.2$^{\circ}$S and 19:00 MLT.
During these three days, the average geomagnetic indices were
Kp: 2.6, 4.8, 4.8, PC: 1.8, 3.9, 4.4, Ap: 12.9, 54.6, 43.5,
and an approximately constant
$\overline{\text{F}_\text{10.7}}$ of 79
($\log(\overline{\text{F}_\text{10.7}}) \approx$ 4.4).

We find that the variability driven by the geomagnetic indices
captures most of the activity in that particular case;
as expected the model does best at medium activity.
However, our model underestimates the peak magnitude of
the ionization rates during this period,
by about a factor of 2 on the lower activity day (05 April 2010),
and by about a factor of 3 to 4 on the day of largest activity.
For comparison we also show electron ionization rates from
the Atmospheric Ionization during Substorm Model (AISstorm).
AISstorm is based on AIMOS~\cite{Wissing2009},
augmented by several improvements, mainly a geomagnetic grid, substorm dependence,
and a particle specific polar cap size~\cite{Yakovchuk2023}.
AISstorm uses 18 years of POES and Metop satellite data (2001--2018),
categorized by Kp level and substorm activity,
and by geomagnetic APEX location~\cite{Richmond1995} and
MLT with up to 1$^{\circ}$ latitude by 3.75$^{\circ}$ longitude resolution.
Typical average flow maps are presented in~\citeA{Yakovchuk2019}.
Here the AISstorm ionization rates are averaged to 2\,h MLT resolution
to match our model.
We find that the ionization rates from AISstorm differ from
our model with peak rates of about a factor of 5 lower.
They also peak at higher altitudes,
about 110\,km compared to around 105\,km from our model.
While the comparison in Fig.~\ref{fig:profs} shows single profiles at a particular location and time,
a more statistical comparison shows that
the peak ionization rates of both models are rather similar (not shown here).

\subsection{Discussion}\label{sec:discussion}

The fit and residuals shown in Figs.~\ref{fig:fit.ts.NH} and~\ref{fig:res.ts.NH}
show some remaining variability on top of white noise.
However, adding (semi-)annual harmonics or more proxies to the regressors
did not improve the fit quality in a way that would have justified
the increase in number of model parameters.

The choice to include both Kp and Ap comes about because both capture
geomagnetic activity with different scales.
Kp is quasi-logarithmic, and Ap quasi-linear.
Adding yet another activity index, PC, has the advantage of
capturing that activity with a higher time resolution,
1-h (PC) on top of the 3-h variability from Kp and Ap.
The general 11-year solar cycle activity is included via
the 81-day averaged 10.7\,cm radio flux, effectively sampled daily.

The DMSP/SSUSI orbits are sun-synchronous which means they sample
approximately the same latitude/MLT region every orbit,
leading to worse sampling in certain areas, see Fig.~\ref{fig:coeff.npts}.
We try to mitigate this low sampling by limiting the number of proxies
as well as the geomagnetic latitude range of the model.
However, in some of the bins the number of usable data points is still small,
and the fit resulted in ambiguous coefficients.
In those cases, we use bilinear interpolation from the surrounding grid cells.
Note that auroral activity and therefore the atmospheric ionization
is typically constrained to a particular region,
and interpolating or extrapolating to regions
without enough data points might result in non-optimal predictions.

The choice to cut off very low ionization rates is
a purely subjective one, and the exact value might be subject to
further fine tuning.
Similarly, higher resolution in MLT and geomagnetic latitude is also
possible, but both would require reprocessing of the data.
Higher vertical resolution could be obtained by interpolating
the 5-km $\log(q)$ values supplied by the model,
along with cubic-spline interpolation of $\log(q)$ to a finer grid.

We use the energy range as given in the SSUSI data files,
which is limited to 2--20\,keV because of the method used~\cite<see, e.g.>{Knight2018}.
We use a single Maxwellian or Gaussian spectrum based on~\citeA{Bender2021},
and no special attempt is made to include different spectral shapes,
e.g.\ including a high-energy tail~\cite<see, e.g.>{Strickland1993}.
This may underestimate the true ionization rates at altitudes below about 95\,km slightly,
but as has been shown, compared to EISCAT they are still within the range of variability~\cite{Bender2021}.

The validation against ground-based measurements is naturally limited in location.
In our model we extend these results globally,
there are no transition MLTs and we use the same spectra for all latitudes.
Since the comparison study covers a wide range of magnetic local times,
separating the pre- and post-midnight sectors seems justified.
However, the extension to all latitudes is an assumption that might skew the
predictions in cases it is violated.

\section{Conclusions}\label{sec:conclusions}

We present an empirical model of auroral ionization rates
derived from 12 (10) years of DMSP/SSUSI F17 (F18) observations.
The ionization rate profiles are derived from the average electron energy
and energy flux used as input to the ionization rate parametrization
by~\cite{Fang2010}, using NRLMSISE-00~\cite{Picone2002/12/24}
as the background neutral atmosphere and the energy spectra as determined
by our validation study~\cite{Bender2021}.
The model is based on the time series of these profiles
binned in 2-h MLT and 3.6$^{\circ}$ geomagnetic latitude and
calculated at 5-km altitude steps from 90 to 150 km.
For each altitude we derived a map of regression coefficients
for Kp, PC, Ap, $\log(\overline{\text{F}_\text{10.7}})$,
and a constant offset.

The choice of proxies has been empirically determined,
taking also into account the availability and how well
the indices are maintained.
When setting up the model, AE and Dst were other
possible choices.
However, data for both indices were only available until
February 2018 and not suitable to fit the whole time series,
even less so for predicting ionization rates for chemistry--climate models.
Of course this caution of index availability also applies to the ones used in this study,
especially the more rarely used PC index.
Currently, the OMNI database is well-maintained in that regard,
which gives us confidence that the choice we made here
is also future proof.
On the other hand, Kp accounted for the majority of the variability,
and it is one of the widest used and most accessible geomagnetic indices
so that this would be the index of choice if the
number of proxies needs to be limited further.

%Text here ===>>>

\appendix
\section{Model application}\label{sec:model.use}

The application of the model to retrieve the ionization rates
should be rather straight-forward:
\begin{enumerate}
  \item Acquire the 1-h sampled proxy values for the UT time in question,
    for example from the OMNI database~\cite{Papitashvili2020}.
  \item If not already available, calculate the geomagnetic latitude
    and magnetic local time for your location and time.
  \item\label{algo:coeff}
    Retrieve the proxy coefficients from the coefficient table file
    for the grid cell containing the location calculated in the previous step,
    i.e.\ the closest grid cell centre as given in the table.
    If preferred, bilinear interpolation in magnetic local time and geomagnetic
    latitude can be used to calculate the coefficients and their variances for
    the location in question.
    Optionally, retrieve the respective coefficients for all
    5-km altitude steps for later interpolation.
  \item\label{algo:mult} Multiply the proxy coefficients by the proxy values.
  \item\label{algo:sum}  Sum over the proxies and add the offset to obtain
    (the logarithm of) the ionization rate at the required altitude.
    Steps~\ref{algo:mult} and~\ref{algo:sum} are the practical implementation
    of Eq.~\eqref{eq:regressmodel}.
  \item (optional) Use cubic (3rd-order) spline interpolation of $\log(q)$
    in altitude to obtain a finer resolution of the ionization rate profile.
  \item (optional) Repeat steps~\ref{algo:coeff} to~\ref{algo:sum},
    replace the proxy coefficients by the ones for the standard deviation
    of the coefficients, and square the results before summing the terms,
    to obtain an estimate for the (squared) uncertainty of $\log(q)$.
\end{enumerate}

\section*{Open Research Section}
% \note[Wiley]{%
% This section MUST contain a statement that describes where the data
% supporting the conclusions can be obtained. Data cannot be listed as
% ``Available from authors'' or stored solely in supporting information.
% Citations to archived data should be included in your reference list.
% Wiley will publish it as a separate section on the paper's page.
% Examples and complete information are here:
% \url{https://www.agu.org/Publish-with-AGU/Publish/Author-Resources/Data-and-Software-for-Authors}
% }
The SSUSI data used in this study are available at
\url{https://ssusi.jhuapl.edu/data_products}~\cite{SSUSI2020}.
The source code to calculate the ionization rates and the empirical model
is available as a Python software package distributed
under the GPLv2 license~\cite{Bender2023a}.

\acknowledgments
Stefan Bender and Patrick J.\ Espy acknowledge support from the Birkeland Center
for Space Sciences (BCSS), supported by the Research Council of
Norway under grant number 223252/F50.
Stefan Bender also acknowledges financial support from the Agencia Estatal de Investigacion,
MCIN/AEI/10.13039/501100011033, through grants PID2022-141216NB-I00 and CEX2021-001131-S.
Larry J.\ Paxton is the principal investigator of the SSUSI project,
funded by
the Air Force Defense Meteorological Satellite Program contract N00024-22-D-6404.
We acknowledge use of NASA/GSFC's Space Physics Data Facility's OMNIWeb service and OMNI data.
The computations were performed on resources provided by
UNINETT Sigma2 - the National Infrastructure for
High Performance Computing and Data Storage in Norway.
We thank J.\ M.\ Wissing for providing the AISstorm data,
the development of AISstorm was funded by the German Science Foundation
(DFG project WI4417/2-1).

%%%%%%%%%%%%%%%%%%%%%%%%%%%%%%%%%%%%%%%%%%%%%%%
% REFERENCES and BIBLIOGRAPHY
%
% \bibliography{<name of your .bib file>} don't specify the file extension
% don't specify bibliographystyle
%
%%%%%%%%%%%%%%%%%%%%%%%%%%%%%%%%%%%%%%%%%%%%%%%

\bibliography{bigbib}

%Reference citation instructions and examples:
%
% Please use ONLY \cite and \citeA for reference citations.
% \cite for parenthetical references
% ...as shown in recent studies (Simpson et al., 2019)
% \citeA for in-text citations
% ...Simpson et al. (2019) have shown...
%
%
%...as shown by \citeA{jskilby}.
%...as shown by \citeA{lewin76}, \citeA{carson86}, \citeA{bartoldy02}, and \citeA{rinaldi03}.
%...has been shown \cite{jskilbye}.
%...has been shown \cite{lewin76,carson86,bartoldy02,rinaldi03}.
%... \cite <i.e.>[]{lewin76,carson86,bartoldy02,rinaldi03}.
%...has been shown by \cite <e.g.,>[and others]{lewin76}.
%
% apacite uses < > for prenotes and [ ] for postnotes
% DO NOT use other cite commands (e.g., \citet, \citep, \citeyear, \nocite, \citealp, etc.).
%

\end{document}